# Superconducting Polycrystalline Rhenium Films Deposited at Room Temperature

Serafim Teknowijoyo (https://orchid.org/0000-0002-5542-5482) and
Armen Gulian* (https://orchid.org/0000-0002-4695-4059)
*Advanced Physics Laboratory, Institute for Quantum Studies,
Chapman University, Burtonsville, MD 20866, USA*

*gulian@chapman.edu

**Abstract** – We report on magnetron deposition of thin superconducting rhenium films on sapphire substrates. During the deposition, substrates were held at ambient temperature. Critical temperature of the films is $T_c \sim 3.6K$. Films have polycrystalline structure, and grazing incidence X-ray diffractometry indicates that crystalline lattice parameters are somewhat larger compared to the bulk ones. Magnetoresistive and AC/DC susceptibilities allowed us to determine $H_{c1}$ and $H_{c2}$ of these films, as well as estimate coherence length $\xi(0)$ and magnetic penetration depth $\lambda_L(0)$. We also provide information on surface morphology of these films.

**Introduction**

Non-centrosymmetric superconductors (NCS) with broken time-reversal symmetry (TRS) are of great interest to contemporary basic research in superconductivity [1-3]. Namely, these materials provide the unique opportunity to design simple and scalable superconducting devices with nonreciprocal current control, such as diodes, transistors, quadristors, etc. [4-11]. Application of these materials may abandon the necessity of external magnetic fields and sophisticated nano-patterning (such as reported in [12]). Re-based compounds, such as Re-Nb [13], Re-Ti [14], Re-Hf [15] and Re-Zr [16] are of great interest, since muon spectroscopy revealed broken time-reversal symmetry in these materials, which are also non-centrosymmetric. Understanding the properties of pure Re films is crucial for further progress with these materials, as well as for the discovery of new compounds. Additionally, there were reports on very large perpendicular critical fields in Re films quenched condensed onto liquid-nitrogen cooled substrates [17]. This continues substantial interest in liquid-helium temperature superconductors which can be used in quantum information and computation technologies [18–20]. Thin-film superconductors can be easily deposited, and they are resistant to oxidation, have low resistivity, and/or are compatible with high magnetic fields [21,22]. Also, Re can be an attractive alternative to Al in devices such as superconducting resonators and microwave circuits [23].

When Re films are deposited onto $850^0C$ substrates, one obtains the bulk $T_c$ [24] which is about 1.7K [25]. Much better results ($T_c \sim 2-6K$) are obtainable at the deposition of nanoscale thickness films on substrates that are cooled to liquid nitrogen level [26-28], as well as via electroplating [29]. In this work we report on successful deposition of polycrystalline Re films on substrates at *ambient temperature* yielding $T_c \sim 3.6K$ and greatly simplifying the deposition route. The presented results on the crystalline structure of the films may provide a key to quantitative description of the mechanism of superconductivity in these materials. We also studied the morphology of the films, magnetotransport, critical fields and magnetic susceptibility which allowed us to estimate basic features of superconducting state, such as the critical fields, coherence length, and London penetration depth.

## 2. Sample preparation

Rhenium films for our study were prepared in conventional manner using DC magnetron option of our multifunctional deposition system (manufactured by AJA International, Inc.) with a base pressure of $1 \times 10^{-8}$ Torr (cryogenic vacuum). The Re target (ACI Alloys, Inc., 99.99% purity) was accommodated inside of a 1.5″ DC gun. The sapphire substrates (University Wafers, thickness 430 μm, C-cut, single-side polished) were cleaned thoroughly with isopropyl alcohol before being mounted on the rotatable substrate holder. In our chamber's configuration, the substrate holder is at the center of the chamber facing upwards, while the five sputtering guns are located at the top. The substrate was rotated in-plane throughout the whole deposition process to ensure homogeneous deposition layer over the whole surface. Our pre-deposition in-situ cleaning of substrates involves a gentle bombardment by Ar+ ions using the Kauffman source in addition to thermal cleaning via heating.

Two batches of samples were prepared: i) deposition of Re on heated substrates and ii) deposition of Re on cold substrates. Corresponding preparation routes were a bit different.

In the hot-substrate Case (i), the Kauffman source ion bombarded perpendicular to the substrate surface for 1 min while the substrates were at $600^0$C. Then temperature was raised to $900^0$C and kept at that temperature for 30 min. Afterwards, the temperature was reduced to $600^0$C and the deposition took place for 10 min, at pressure 4 mTorr, with gun power 250W, and anode voltage 465 V. After the deposition, the temperature was raised back to $900^0$C for in-situ annealing for 30 min and then cooled down to ambient temperature. All the heating/cooling protocols consistently used a 30°C/min ramp rate.

In the cold-substrate Case (ii), the Kauffman source bombarded at 45 degrees to the substrate surface for the same 1 minute duration. Substrates were not heated. Deposition, as above, took 10 min, at argon pressure 3 mTorr (first 5 min) and 4 mTorr (last 5 min). Gun power was again 250 W, and the anode voltage was ~540 V (first 5 min) and ~500 V (last 5 min). No annealing was applied.

## 3. Experimental results

Our films were ~100 nm thick, as was established via reflective X-ray interferometry using Rigaku SmartLab X-ray diffraction apparatus (inset to Fig. 1).

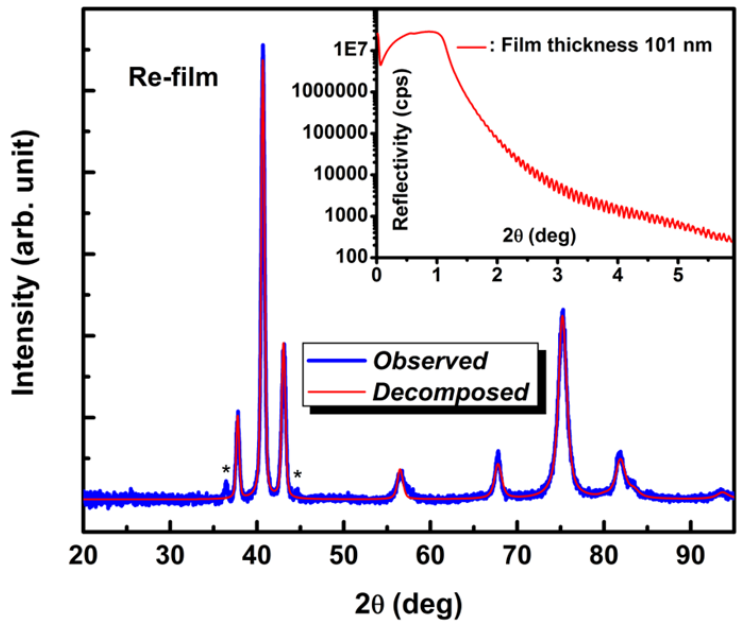

**Fig. 1.** XRD obtained via grazing incidence method indicates the crystalline structure of the "cold-substrate" film (this method excludes substrate reflections). Its reflectivity curve (inset) is used for determining the film thickness. Theoretical fit is obtained via the Decomposition method (data are in the main text). Asterisks indicate small unidentified peaks.

The films of the "hot-substrate" batch were amorphous and down to 1.8K no superconducting phase was noticed. As mentioned by other researchers, the films deposited at higher temperatures demonstrate bulk Re superconducting phase with $T_c \approx 1.7K$ which is also most likely our case (we did not cool down the films below 1.8K). At the same time, the resistivity of these amorphous films at 300K is a factor of three lower than that of the superconducting batch deposited on cold substrates: $\rho^{300K}_{cold} \approx 12$ µΩ cm. The x-ray data corresponding to polycrystalline phase films (Case ii) are shown in Table 1.

**Table 1. Crystal structure comparison**

| Lattice parameters | a, Å | b, Å | c, Å | volume, Å$^3$ | space group |
|---|---|---|---|---|---|
| **Our films** | 2.782 | 2.782 | 4.484 | 30,053 | P63/mmc |
| **Bulk Re** | 2.761 | 2.761 | 4.458 | 29.430 | P63/mmc |

Our data shown in this table are obtained using the XRD curve in Fig. 1 for polycrystalline superconducting batch corresponding to Case (ii). Special grazing incidence method of analysis of Rigaku SmartLab system was used which excludes the appearance of substrate reflections. The rest of this article is focused on the superconducting batch of this Case (ii).

Figure 2 demonstrates surface morphology of superconductor films.

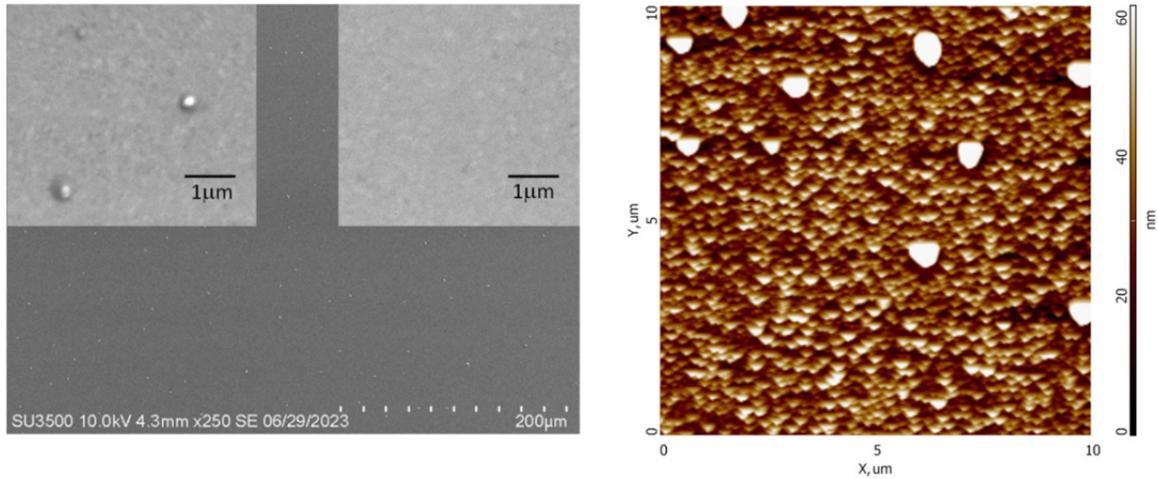

**Fig. 2.** Surface morphology of Re films deposited without heat treatment. Left panel: SEM image (Hitachi SU3500 scanning electron microscope) with two zoomed-in insets. Right panel: AFM image of the same sample. Granular structure is visible with both methods.

Superconducting transition of our films are shown in Fig. 3 (DC resistivity of a sample acquired with three preferential directions) and Fig. 4 (magnetic properties). These data were all acquired by the Quantum Design DynaCool PPMS system. Figure 4 contains real and imaginary parts of AC susceptibility, panels (a) and (b); ZFC and FC patterns of M(H), panel (c) and classic butterfly feature, panel (d), as well as its virgin curves, panel (e), all obtained with DC moment measurement; the last panel (f) constructs the value of $H_{c1}$ using the data of panel (e) via modeling in accordance with the relation $H_{c1}(T)=H_{c1}(0)[1-(T/T_c)^2]$. The values of transition temperature which follow from resistive and magnetic measurements are in accordance with each other yielding $T_c(H=0) \approx 3.57K$.

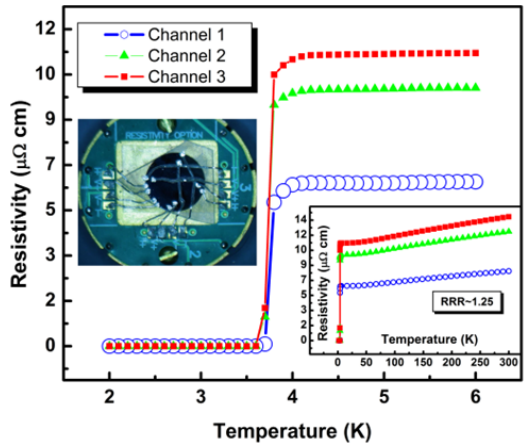

**Fig. 3.** Superconducting transitions in our films documented by DC resistivity at H=0. Left inset shows sample mounted on PPMS puck with electric leads (indium wire) corresponding to Ch. 1-3. By purpose, leads for Ch. l are arranged at the film edge, and for Ch. 2 and Ch. 3 in mutually crossed geometry to examine homogeneity of the superconducting phase.

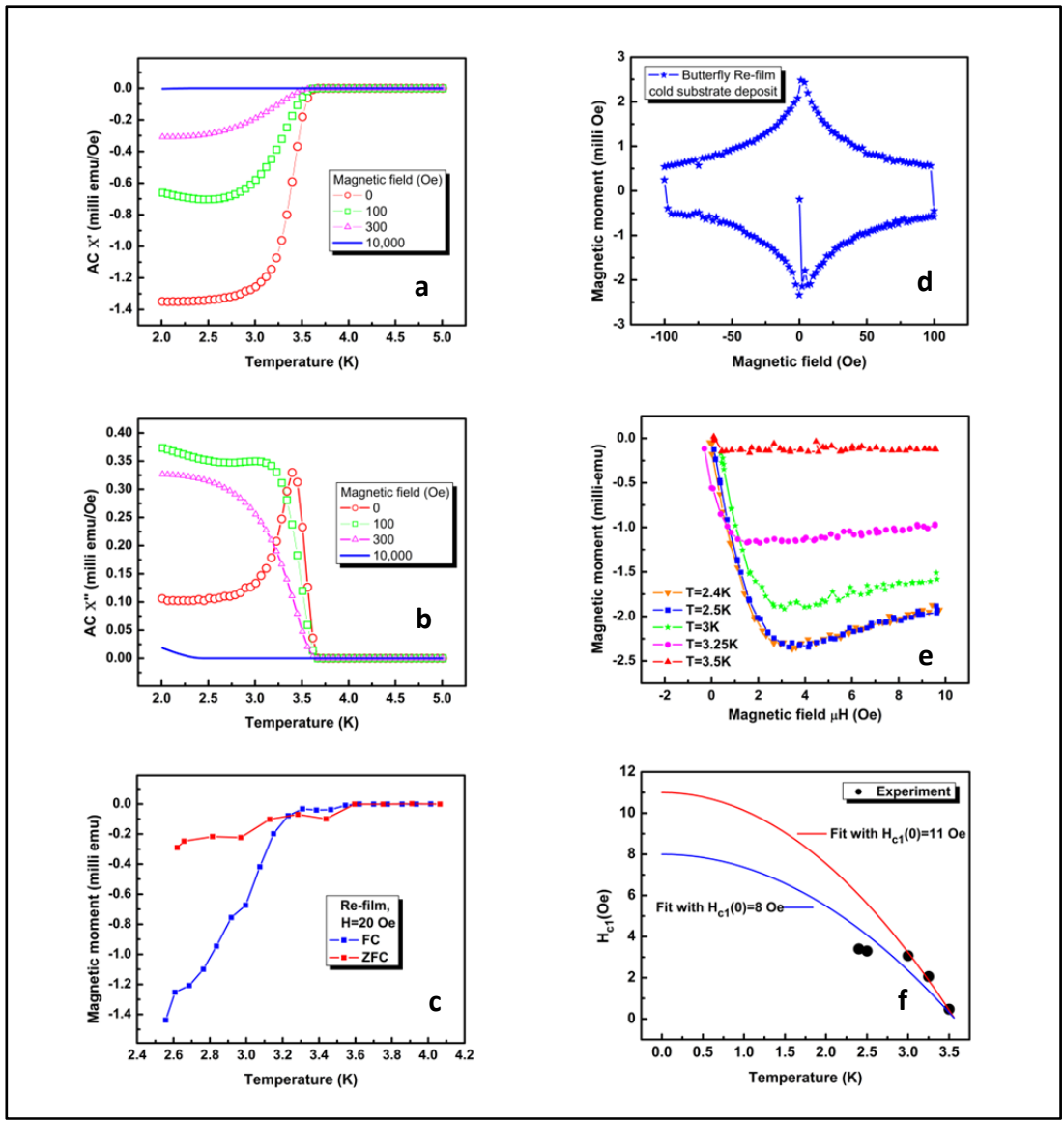

**Fig. 4.** Magnetic characterization of Re films. Panels (a)-(f) are described in the main text. AC measurements are run at frequency 4000Hz, with AC amplitude 2 Oe.

## 4. Discussion

Our sample's $H_{c2}$ (T) curve shows a different behavior compared to the conventional BCS dependence $H_{c2}(T) = H_{c2}(0)[1 - (T/T_c)^2]$. Therefore, the curve was instead fitted using the expression $H_{c2}(T) = H_{c2}(0)[1 - (T/T_c)^p]^q$ following [30,31] where the exponents p = q = 3/2. A slightly better fit to the data can be obtained when the constraint on p and q are removed (Fig. 5).

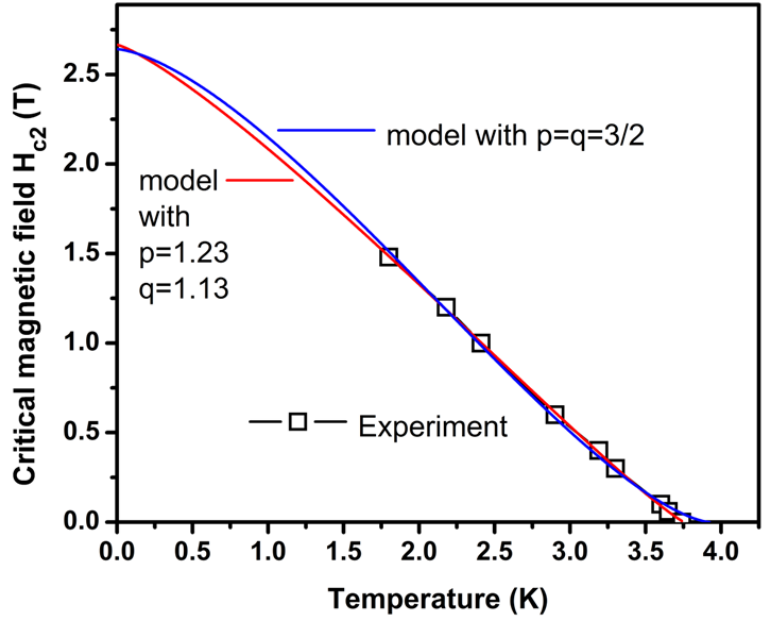

**Fig.5.** Fitting experimental data by customized modeling curves.

Nevertheless, both fitting procedures give a similar value for $H_{c2} \approx 26.5$ kOe. Using the Ginzburg-Landau relation $H_{c2} = \phi_0/(2\pi\xi^2)$, where $\phi_0 = 2.068 \times 10^{-15}$ Wb is the flux quantum [32], the estimated coherence length of our film is $\xi(0) \approx 11$ nm. Combining $\xi$ and the estimate for $H_{c1}$ from panel (f) in Fig. 4 ($H_{c1} \sim$8-11 Oe) into the standard expression $H_{c1} = [\phi_0/(4\pi\lambda^2)][\ln(\lambda/\xi) + 0.12]$ [33], the estimate for our film's magnetic penetration depth is $\lambda(0) \approx 9.8$ nm. Finally, the Ginzburg-Landau parameter can also be calculated, $\kappa = \lambda/\xi = 0.89$, which shows that our film, being of type-II, is rather close to the theoretical boundary separating type-I and type-II superconductors of $\kappa = 1/\sqrt{2} = 0.71$.

It is interesting to notice that our room-temperature resistivity ($\rho_{film} \approx 12 \mu\Omega$ cm) is somewhat smaller than $\rho_{bulk} \approx 18 \mu\Omega$ cm of the bulk rhenium [27]. Even more interesting is the fact that the lattice parameters of our films when compared with the bulk sample parameters (see Table 1) indicate expansion. Meanwhile, those obtained by Frieberthauser and Notarys [28]: *a=b*=2.748Å and *c*=4.44Å indicate compression. One interesting remark by these authors (who addressed the influence of film thickness on superconducting $T_c$) is related with the fact that 1000 Å films (which is our case) have the lowest $T_c$ (lower than 3K in their case), and thicker or thinner films reveal higher values (>4K). From that point of view one may expect higher transition temperatures for thinner or thicker films using our approach.

## 5. Conclusions

Information obtained by our research may provide grounds for further fundamental studies based on band-structure computations of superconducting state in these materials to quantitatively explain the more than two-fold increase in $T_c$ compared to the bulk. Also, the simplicity of the deposition method may be feasible for wide range of device applications mentioned in Introduction. Finally, the parameters $\lambda_L$ and $\xi$ estimated above may be used for modeling of phenomena in these devices.

*Acknowledgements*
We are indebted to Dr. E. Vinogradova (Rigaku USA) for her help with processing the XRD data. We are grateful to the Physics Art Frontiers for the provided technical assistance. This research was supported by the ONR grants No. N00014-21-1-2879 and No. N00014-20-1-2442.

## References

1. Y. Tokura and N. Nagaosa, Nonreciprocal responses from non-centrosymmetric quantum materials, *Nature Commun.* **9**, 3740 (2018).
2. R. Wakatsuki and N. Nagaosa, Nonreciprocal Current in Noncentrosymmetric Rashba Superconductors, *Phys. Rev. Lett.* **121**, 026601 (2018).
3. S. Hoshino, R. Wakatsuki, K. Hamamoto, and N. Nagaosa, Nonreciprocal charge transport in two dimensional noncentrosymmetric superconductors, *Phys. Rev. B* **98**, 054510 (2018).
4. F. Ando, Y. Miyasaka, T. Li, J. Ishizuka, T. Arakawa, Y. Shiota, T. Moriyama, Y. Yanase, and T. Ono, Observation of superconducting diode effect, *Nature* **584**, 373 (2020).
5. T. Ideue and Y. Iwasa, One-way supercurrent achieved in an electrically polar film, *Nature* **584**, 349 (2020).
6. C. Baumgartner, L. Fuchs, A. Costa, S. Reinhardt, S. Gronin, G. C. Gardner, T. Lindemann, M. J. Manfra, P. E. Faria Junior, D. Kochan, J. Fabian, N. Paradiso, and C. Strunk, Supercurrent rectification and magnetochiral effects in symmetric Josephson junctions, *Nature Nanotechnology* **17**, 39 (2022).
7. H. Wu, Y. Wang, Y. Xu, P. K. Sivakumar, C. Pasco, U. Filippozzi, S. S. P. Parkin, Y.-J. Zeng, T. McQueen, and M. N. Ali, The field-free Josephson diode in a van der Waals heterostructure, *Nature* **604**, 653 (2022).
8. E. Strambini, M. Spies, N. Ligato, S. Ilic, M. Rouco, C. Gonzalez-Orellana, M. Ilyn, C. Rogero, F. S. Bergeret, J. S. Moodera, P. Virtanen, T. T. Heikkila, and F. Giazotto, Superconducting spintronic tunnel diode, *Nature Commun.* **13**, 2431 (2022).
9. T. Morimoto and N. Nagaosa, Nonreciprocal current from electron interactions in noncentrosymmetric crystals: roles of time reversal symmetry and dissipation, *Scientific Reports* **8**, 2973 (2018).
10. S. Chahid, S. Teknowijoyo, I. Mowgood, and A. Gulian, High-frequency diode effect in superconducting $Nb_3Sn$ microbridges, *Phys. Rev. B* **107**, 054506 (2023).
11. S. Teknowijoyo, S. Chahid, and A. Gulian, Flux quanta injection for nonreciprocal current control in a 2D noncentrosymmetric superconducting structure, *Phys. Rev. Applied* (accepted).
12. Y.-Y. Lyu, J. Jiang, Y.-L. Wang, Z.-L. Xiao, S. Dong, Q.-H. Chen, M. V. Milošević´, H. Wang, R. Divan, J. E. Pearson, P. Wu, F.M. Peeters, and W.-K. Kwok, Superconducting diode effect via conformal-mapped nanoholes, *Nat. Commun.* **12**, 2703 (2021).

13. T. Shang, M. Smidman, S. K. Ghosh, C. Baines, L. J. Chang, D. J. Gawryluk, J. A. T. Barker, R. P. Singh, D. McK. Paul, G. Balakrishnan, E. Pomjakushina, M. Shi, M. Medarde, A. D. Hillier, H. Q. Yuan, J. Quintanilla, J. Mesot, and T. Shiroka, Time-Reversal Symmetry Breaking in Re-Based Superconductors, *Phys. Rev. Lett.* **121**, 257002 (2018).
14. D. Singh, Sajilesh K. P., J. A. T. Barker, D. McK. Paul, A. D. Hillier, and R. P. Singh Time-reversal symmetry breaking in the noncentrosymmetric superconductor $Re_6Ti$, *Phys. Rev. B* **97**, 100505(R) (2018).
15. D. Singh, A. D. Hillier, A. Thamizhavel, and R. P. Singh, Superconducting properties of the noncentrosymmetric superconductor $Re_6Hf$, *Phys. Rev. B* **94**, 054515 (2016).
16. S. Dutta, V. Bagwe, G. Chaurasiya, A. Thamizhavel, R. Bapat, P. Raychaudhuri, S. Bose, Superconductivity in amorphous $Re_xZr$ (x≈6) thin films, *J. Alloys and Compounds*, **877**, 160258 (2021).
17. F. N. Womack, D. P. Young, D. A. Browne, G. Catelani, J. Jiang, E. I. Meletis, and P. W. Adams, Extreme high-field superconductivity in thin Re films, *Phys. Rev. B* **103**, 024504 (2021).
18. T. M. Hazard, A. Gyenis, A. Di Paolo, A. T. Asfaw, S. A. Lyon, A. Blais, and A. A. Houck, Nanowire Superinductance Fluxonium Qubit, *Phys. Rev. Lett.* **122**, 010504 (2019).
19. L. Grünhaupt, M. Spiecker, D. Gusenkova, N. Maleeva, S. T. Skacel, I. Takmakov, F. Valenti, P. Winkel, H. Rotzinger, W. Wernsdorfer, A. V. Ustinov, and I. M. Pop, Granular aluminium as a superconducting material for high-impedance quantum circuits, *Nat. Mater.* **18**, 816 (2019).
20. D. Niepce, J. Burnett, and J. Bylander, High Kinetic Inductance NbN Nanowire Superinductors, *Phys. Rev. Appl.* **11**, 044014 (2019).
21. N. Samkharadze, A. Bruno, P. Scarlino, G. Zheng, D. P. DiVincenzo, L. DiCarlo, and L. M. K. Vandersypen, High-Kinetic-Inductance Superconducting Nanowire Resonators for Circuit QED in a Magnetic Field, *Phys. Rev. Appl.* **5**, 044004 (2016).
22. K. Borisov, D. Rieger, P. Winkel, F. Henriques, F. Valenti, A. Ionita, M. Wessbecher, M. Spiecker, D. Gusenkova, I. M. Pop, and W. Wernsdorfer, Superconducting granular aluminum resonators resilient to magnetic fields up to 1 Tesla, *Appl. Phys. Lett.* **117**, 120502 (2020).
23. E. Dumur, B. Delsol, T. Weissl, B. Kung, W. Guichard, C. Hoarau, C. Naud, K. Hasselbach, O. Buisson, K. Ratter, and B. Gilles, Epitaxial Rhenium Microwave Resonators, *IEEE Trans. Appl. Supercond.* **26**, 1501304 (2016).
24. C. Song, T. W. Heitmann, M. P. DeFeo, K. Yu, R. McDermott, M. Neeley, John M. Martinis, and B. L. T. Plourde, Microwave response of vortices in superconducting thin films of Re and Al, *Phys. Rev. B* **79**, 174512 (2009).
25. B.W. Roberts, Survey of Superconducting materials and Critical Evaluation of Selected Properties, *J. Phys. Chem. Ref. Data*, **5**, 581 (1976).
26. N. E. Alekseevskii, M. N. Mikheeva, N. A. Tulina, The superconducting properties of rhenium, *Zh. Eksp. Teor. Fiz.* **52**, 875(1967) [*Sov. Phys. JETP* **25**, 575 (1967)].
27. A. Ul Haq and O. Meyer, Electrical and superconducting properties of rhenium thin films, *Thin Solid Films*, **94** 119 (1982).
28. P. E. Frieberthauser and H. A. Notarys, Electrical properties and superconductivity of rhenium and molybdenum films, *J. Vacuum Sci. and Technol.*, **7**, 485 (1970).
29. D. P. Pappas, D. E. David, R. E. Lake, M. Bal, R. B. Goldfarb, D. A. Hite, E. Kim,

H.-S. Ku, J. L. Long, C. R. H. McRae, L. D. Pappas, A. Roshko, J. G. Wen, B. L. T. Plourde, I. Arslan, and X. Wu, Enhanced superconducting transition temperature in electroplated rhenium, *Appl. Phys. Lett.* **112**, 182601 (2018); https://doi.org/10.1063/1.5027104.
30. P. K. Biswas, M. R. Lees, A. D. Hillier, R. I. Smith, W. G. Marshall, and D. McK. Paul, Structure and superconductivity of two different phases of $Re_3W$, *Phys. Rev.* B **84**, 184529 (2011).
31. R. Micnas, J. Ranninger, and S. Robaszkiewicz, Superconductivity in narrow-band systems with local nonretarded attractive interactions, *Rev. Mod. Phys.* **62**, 113 (1990).
32. E. H. Brandt, Flux distribution and penetration depth measured by muon spin rotation in high-$T_c$ superconductors, *Phys. Rev. B* **37**, 2349 (1988).
33. M. Tinkham, Introduction to Superconductivity (McGraw-Hill, New York, 1975).